\documentclass[12pt]{iopart}

\usepackage{graphicx}% Include figure files
\usepackage{epstopdf}
\usepackage{dcolumn}% Align table columns on decimal point
\usepackage{bm}% bold math
\usepackage{tabularx}
\usepackage{multirow}

\begin{document}

\title[Isotope shifts and hyperfine structure measurements of tungsten]{High resolution isotope shifts and hyperfine structure measurements of tungsten by laser induced fluorescence spectroscopy}

\author{Jeongwon Lee, Jinhai Chen, and Aaron Leanhardt}

\address{
Department of Physics, University of Michigan, Ann Arbor, Michigan 48109-1040, USA
}

\date{\today}

\begin{abstract}
Isotope shifts and hyperfine structure of tungsten were studied in the near UV range. We have used laser induced fluorescence spectroscopy on a pulsed supersonic beam to probe the $^5D_0$ $\rightarrow$ $^5F_1$ transition at 384.9 nm, $^7S_3$ $\rightarrow$ $^7P_4$ transition at 400.9 nm, and $^7S_3$ $\rightarrow$ $^7P_3$ transition at 407.4 nm. Three new magnetic hyperfine constants are reported for $^7P_3$,$^7P_4$, and $^5F_1$ states. The isotope shifts of the 384.9 nm transition are presented for the first time, and the isotope shifts of 400.9 nm and 407.4 nm transition are measured with an order of magnitude higher precision compared to the previous measurements. As a result, the nuclear parameters $\lambda$ and $\lambda_{rel}$ are extracted from the isotope shifts with an improved precision.
\end{abstract}

\maketitle

\section{\label{sec:level1}Introduction}

High resolution spectroscopic study of tungsten(W) is of interest in various fields of physics~\cite{Morton2000,Buttgenbach1982,Aufmuth1988a,Aufmuth1988b,Neu1996,Federici2001}. As one of 5d shell atoms, its isotope shifts and hyperfine structure allow enhanced understanding of nuclear structure in the deformed region on the nuclear chart. B\"{u}ttgenbach~\cite{Buttgenbach1982} analyzed the hyperfine structure of 4d- and 5d-shell atoms to study contact interaction terms. Aufmuth~\cite{Aufmuth1988b} extracted nuclear parameter $\lambda$ from tungsten optical isotope shifts.

More recently there has been great interest in using tungsten as plasma facing material in controlled fusion~\cite{Neu1996}. Particularly, the International Thermonuclear Experimental Reactor (ITER) plans to use W tiles on the divertor plasma~\cite{Federici2001}, along with the spectroscopic diagnostic system~\cite{Sugie2005} to estimate the tungsten influx rate for plasma edge modeling. According to Skinner~\cite{Skinner2009}, the $^7S_3$ $\rightarrow$ $^7P_4$ transition of neutral tungsten at 400.9 nm is considered promising due to its high transition probability. However, there is a complication in spectroscopic diagnostics due to singly ionized tungsten lines(W II) coincidentally at nearly the same wavelength as the neutral lines(W I), causing line blending issues~\cite{Skinner2009}. Another problem is the incomplete information on $^7S_3$ $\rightarrow$ $^7P_4$ transition, as the latest study did not reveal the hyperfine structure due to limited resolution~\cite{Aufmuth1988b}. This would lead to further difficulties in accurate modeling of the tungsten influx rate at ITER.

In this paper, we present isotope shifts and hyperfine structure measurements of three neutral tungsten transitions in the near UV range. Laser Induced Fluorescence (LIF) spectroscopy on an atomic beam were performed on $^5D_0$ $\rightarrow$ $^5F_1$ transition at 384.9 nm, $^7S_3$ $\rightarrow$ $^7P_4$ transition at 400.9 nm and, $^7S_3$ $\rightarrow$ $^7P_3$ transition at 407.4 nm. Magnetic dipole hyperfine constant of $^7S_3$ level is compared with the prior report~\cite{Buttgenbach1979}, and three new measurements are made for the constants of $^7P_3$, $^7P_4$ and $^5F_1$ levels.

The isotope shifts of 384.9 nm transition is presented for the first time and the isotope shifts of 400.9 nm and 407.4 nm were measured with approximately 10 times higher precision compared with the previous results. High resolution isotope shift study of these three transitions enabled us to extract the nuclear parameters $\lambda$ and $\lambda_{rel}$, which were compared with the previous results~\cite{Aufmuth1988b}. Also, completely resolved tungsten spectrum at 400.9 nm is directly related to the application at ITER~\cite{Skinner2009}. Experimental details are shown in section 2 and 3, followed by analysis in section 4,5 and conclusion in section 6.

\section{Experimental Methods}

The beam production of refractory metals has been limited due to its high melting points. Having the highest melting point of all metals, tungsten atomic beam cannot be generated efficiently through conventional oven method~\cite{Scoles1988}. In the past, we have used a resistively heated tungsten wire to generate a high flux beam~\cite{Lee2009}. However, the Blackbody radiation coming from the heated wire was limiting our ability to detect fluorescing light from the atomic transitions. Here, we use Smalley type pulse supersonic beam technique~\cite{Dietz1981} to overcome these issues. Only a brief review of the technique is given here and the details can be found in ref.~\cite{Dietz1981}.

\begin{figure}
\includegraphics[width = 3.3 in]{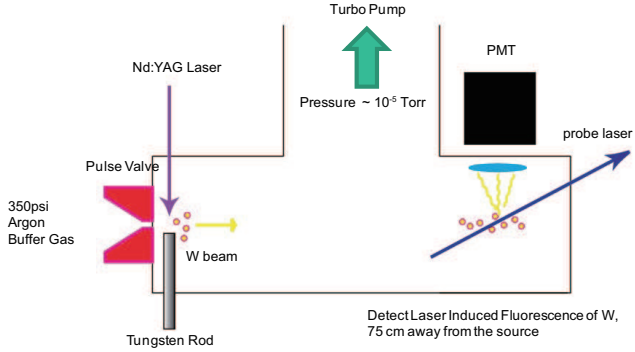}% Here is how to import EPS art
\caption{Diagram of Tungsten Beam Apparatus}
\label{f:Vacuum}
\end{figure}

Tungsten atoms are ablated from a rod (American Elements, $99.9\%$ purity) by the third harmonic of the Nd:YAG pulse laser (Quantel), while the solenoid gas valve (Parker, general valve series 999) entrains the atoms with Argon buffer gas. The atoms get cooled down through buffer gas collisions resulting narrow Doppler linewidth. The turbo pump with $1500$L/s of pumping speed maintained the operating pressures to be, $5\times 10^{-6}$ Torr inside the vacuum chamber. Diagram of the experimental apparatus is shown in Fig.~\ref{f:Vacuum}.

A tunable cw Ti-SAPH laser(Coherent MBR110) generated the IR light, which was frequency doubled with LBO crystal inside a cavity(Coherent MBD200) to produce the light in 380 nm - 410 nm range. This wide range of tunability enabled us to probe all the transitions presented in this paper using the same laser system. The probe laser was focused with an intensity of $\sim$ $1mW/cm^2$ at the intersection point where the laser beam is crossing the atomic beam perpendicularly. The laser induced fluorescing light was collected by a spherical lens into a water cooled Photo Multiplying Tube(Hammamatsu) connected to the photon counter. The photon counts were recorded simultaneously as the wavelength meter(High Finesse WSU series) measures the frequency of the probe laser with an accuracy level of few MHz. From the methods that are shown in the Appendix section, we have estimated the total uncertainty of frequency measurements of, 1.7MHz for 384.9 nm transition, 2.1MHz for 400.9 nm transition, and 2.9MHz for 407.4 nm transition.

\section{Experimental Results}

\begin{figure}
\includegraphics[width = 3.3 in]{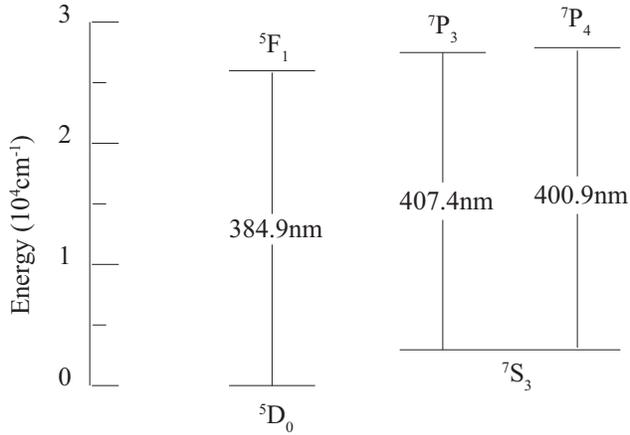}% Here is how to import EPS art
\caption{Tungsten transitions and related energy levels.}
\label{f:Elevel}
\end{figure}

Three of the measured W optical transitions are shown in figure~\ref{f:Elevel} with their relevant energy levels. The 384.9 nm transition is from the $^5D_0$ ground electronic state with the $5d^46s^2$ configuration to the $^5F_1$ excited state with $5d^36s^26p$ configuration~\cite{Wyart2010}. Two other transitions of 400.9 nm and 407.4 nm shares the same low lying metastable state of $^7S_3$ with $5d^56s$ configuration, and excited states of $^7P_4$ and $^7P_3$ both have $5d^46s6p$ configuration~\cite{Wyart2010}.

\begin{figure}
\includegraphics[width = 3.3 in]{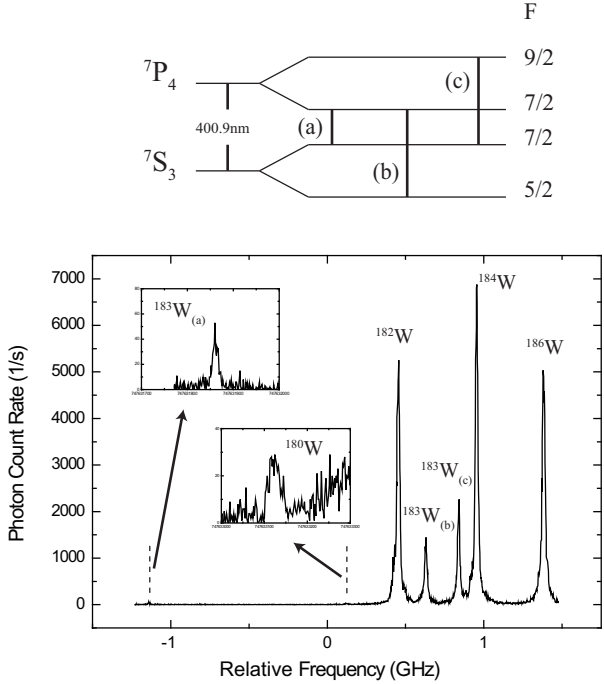}% Here is how to import EPS art
\caption{$^{7}S_3$ $\rightarrow$ $^{7}P_4$ tungsten transition measured by laser induced fluorescence spectroscopy. Dashed lines show the positions of the smaller peaks.}
\label{f:W401line}
\end{figure}

\begin{figure}
\includegraphics[width = 3.3 in]{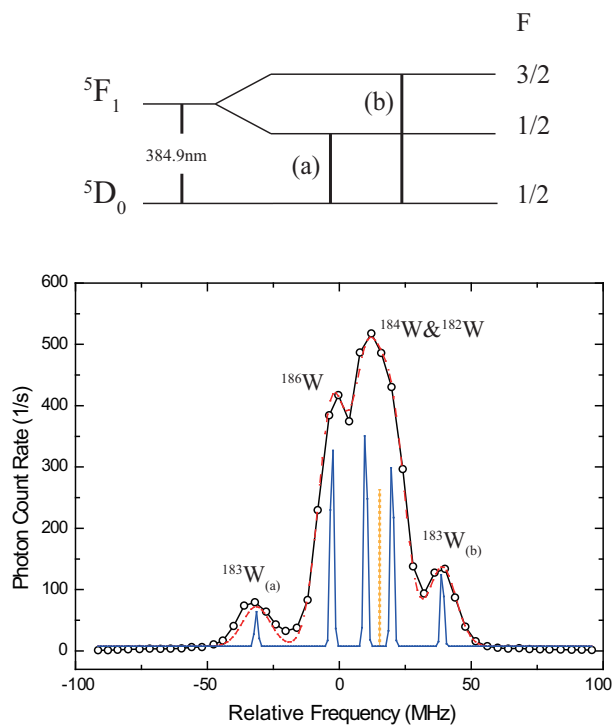}% Here is how to import EPS art
\caption{$^{5}D_0$ $\rightarrow$ $^{5}F_1$ tungsten ground state transition was measured by laser induced fluorescence spectroscopy. Black hollow circles are the experimental data, red dashed line is the least square fit, blue solid line is the simulated line positions, and the yellow dotted line is the center of gravity position for $^{183}$W.}
\label{f:W385line}
\end{figure}

Tungsten has four even isotopes of $^{180}$W,$^{182}$W,$^{184}$W,$^{186}$W, with nuclear spin of $0$, and one odd isotope of $^{183}$W, with nuclear spin of $1/2$ giving rise to hyperfine structure. As shown in Fig.~\ref{f:W401line}$\sim$~\ref{f:W385line}, 400.9 nm, 407.4 nm, and 384.9 nm transitions have three, four, and two allowed electric dipole hyperfine transitions, respectively.

\begin{figure}
\includegraphics[width = 3.3 in]{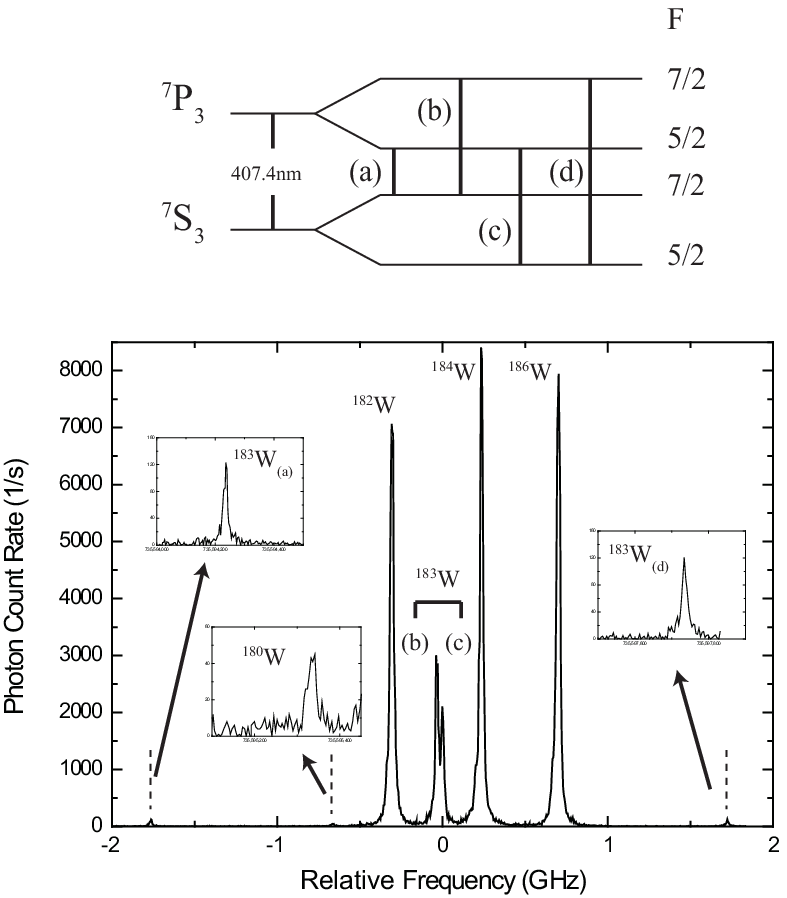}% Here is how to import EPS art
\caption{$^{7}S_3$ $\rightarrow$ $^{7}P_3$ tungsten transition was measured by laser induced fluorescence spectroscopy. Dashed lines show the positions of the smaller peaks.}
\label{f:W407line}
\end{figure}

Previous studies of the 400.9 nm and 407.4 nm transition reported $20\sim30$MHz precision on isotope shift measurements with the hyperfine structures being unresolved~\cite{Aufmuth1988b}. In this paper, we have achieved 10 times higher precision on the isotope shifts measurements of both transitions and the hyperfine structures are clearly resolved as well. These are shown in fig.~\ref{f:W401line} and~\ref{f:W407line}. The ratio of the peak amplitudes agreed within $10\%$ level of the natural abundance ratio combined with the hyperfine line strengths, which was calculated from the Clebsch-Gordan coefficients. The small disagreement comes from the fluctuation of atomic beam intensity. The linewidth of the measurements were mainly limited by Doppler broadening, with a Gaussian linewidth of 5MHz. We extract the line positions of each isotopes from individual Gaussian fits, except for the merged peaks of $^{183}$W$\left(b\right)$ and $^{183}$W$\left(c\right)$ in fig.~\ref{f:W407line}, which we fit to two Gaussians simultaneously. The fit gave an frequency uncertainty of 200kHz in line positions which is negligible compared with the measurement uncertainty described in the previous section.

As seen on figure~\ref{f:W385line}, the isotope shifts of 384.9 nm transition were partly unresolved due to limited resolution, and the $^{180}$W with $0.12\%$ natural abundance is believed to be masked underneath the hyperfine state (b) of $^{183}$W. As some of the peaks weren't resolved, we had to take a different approach to analyze the isotope shifts. First step was to take an average of multiple scans of the transition, in order to eliminate the beam fluctuation effect and get an exact line profile. After this, we used a multi parameter Gaussian fit to extract the isotope and hyperfine shifts. Based on natural abundance ratio and the intensity rule for the hyperfine transitions, we identified two peaks on the wing side in figure~\ref{f:W385line} as the hyperfine states of $^{183}$W. Also, we noticed the center of gravity of $^{183}$W lying closer to the right side of the unresolved center peak, which indicates negative isotope shifts for the 384.9 nm transition. With the above constraints for peak assignments, the least square fit was performed to match the experimental data. The least square fitting curve is shown in red dashed line with a Gaussian linewidth of 5MHz which overlaps well with the experimental data shown in black hollow circles. The blue solid line is the result of the simulation with a linewidth of 0.5MHz to show the line positions, and the yellow dotted line is the center of gravity position of $^{183}$W. The $^{180}$W isotope was excluded from the fit, as the experimental data doesn't have enough resolution to constrain its position. We note that regardless of the peak assignments, the upper limits of $20$ MHz/amu for the isotope shifts and $80$ MHz for the hyperfine splitting of the $^{5}F_1$ state, can be deduced from figure~\ref{f:W385line}. No previous study of isotope shift exists for 384.9 nm transition, however, Gluck~\cite{Gluck1965} has reported 0 isotope shifts for this transition with a frequency resolution of $30$ MHz, which is consistent with our result within their uncertainty of measurement.

\section{W Hyperfine Structure Analysis}

The hyperfine structure of $^{183}$W isotope is only caused by magnetic dipole interaction as it has a nuclear spin of $1/2$~\cite{Kopfermann1958}. In this case, the hyperfine energy levels are given by,

\begin{equation}\label{(1)}
E_{F}=E_{J}+hA\frac{F(F+1)-J(J+1)-I(I+1)}{2},
\end{equation}
where $h$ is the Planck's constant and $A$ is the magnetic hyperfine constant. The shifted energy levels due to this hyperfine interaction are shown on Fig.~\ref{f:W401line}$\sim$~\ref{f:W385line}. The hyperfine frequency splitting within the electronic state becomes,

\begin{equation}\label{(2)}
\delta\nu_{hyperfine}=A\left ( J+\frac{1}{2} \right ).
\end{equation}

It is straightforward to extract the magnetic hyperfine constant $A$ from measured frequency splittings using equation~\ref{(2)}. As some of the hyperfine splittings are measured in multiple transitions, such as the splitting of $^7S_3$ state, we take the weighted mean of the measurements from different transitions to get $A$, which are shown on Table~\ref{table1}. We report three new hyperfine constants for $^7P_3$, $^7P_4$, $^5F_1$ states, and the constant for $^7S_3$ state is compared with the previous result~\cite{Buttgenbach1979}. Good agreement is found within the error of the measurement for the $^7S_3$ state.

Wyart suggested $5d^46s6p$ as the dominant configuration for $^7P_3$, $^7P_4$ states, and $5d^36s^26p$ configuration for the $^5F_1$ state~\cite{Wyart2010}. Our measurements of larger constants $A$ in the cases of $^7P_3$ and $^7P_4$ state compared with the case of $^5F_1$ state, support Wyart's configurations. This is due to the open shell 6s electron from $5d^46s6p$ configuration having bigger contributions to $A$ than the $5d^36s^26p$ configuration.

\begin{table}
\caption{The magnetic dipole hyperfine constant $A$ was measured and compared for $^7S_3$, $^7P_3$, $^7P_4$, and $^5F_1$ electronic states of Tungsten. The 1 $\sigma$ errors are shown inside the parentheses in units of the last decimal quoted.}\label{table1}
\begin{indented}
\lineup
\item[]\begin{tabular}{@{}lll}
\br
&\centre{2}{$A$ (MHz)}\\
\ns
Electronic&\crule{2}\\
States&This work&Ref.~\cite{Buttgenbach1979}\\
\mr
$\0\0\05d^56s$ $^7S_3$&505.5(4)&505.592(12)\\
$\05d^46s6p$ $^7P_3$&496.2(6)&\\
$\05d^46s6p$ $^7P_4$&440.5(6)&\\
$5d^36s^26p$ $^5F_1$&\047.1(11)&\\
\br
\end{tabular}
\end{indented}
\end{table}

\section{W Isotope Shift Analysis}

The theory of atomic isotope shifts are well developed in many publications~\cite{Seltzer1969,Heilig1974,King1984,Aufmuth1987}. Due to the way our experiment is designed, we examine the total isotope shift of the electronic transition rather than the isotope shifts of individual electronic states. The total isotope shift of the electronic transition is equivalent to the difference in isotope shifts of upper and lower electronic states that are involved. Analyzing the isotope shift of individual electronic states will be discuss at the end of this section.

Isotope shift between two isotopes with mass number A and A' of an optical transition $i$  can be written as,

\begin{equation}\label{(3)}
\delta \nu _{i}^{AA'}=\delta \nu _{i,MS}^{AA'}+\delta \nu _{i,FS}^{AA'},
\end{equation}
where MS stands for Mass Shift and FS is for Field Shift.
The Mass shift is further separated into Normal Mass Shift (NMS) and Specific Mass Shift (SMS), both coming from the change in nuclear mass.

\begin{eqnarray}
\delta \nu _{i,MS}^{AA'}&=\delta \nu _{i,NMS}^{AA'}+\delta \nu _{i,SMS}^{AA'}\\
&=\left ( M_{i,NMS}+M_{i,SMS} \right )\frac{A'-A}{AA'}\label{(4)}
\end{eqnarray}

The Field Shift (FS) originates from the change in volume and shape of the nucleus, therefore being directly related to the changes in mean square nuclear charge radii $\delta \left \langle r^{2} \right \rangle$. This relation is shown below,

\begin{equation}\label{(5)}
\delta \nu _{i,FS}^{AA'}=F_{i}\lambda ^{AA'}
\end{equation}

\begin{eqnarray}\label{(6)}
\lambda ^{AA'}&=\delta \left \langle r^{2} \right \rangle^{AA'}+\frac{C_{2}}{C_{1}}\delta \left \langle r^{4} \right \rangle^{AA'}+\frac{C_{3}}{C_{1}}\delta \left \langle r^{6} \right \rangle^{AA'}+\cdots \\
&\approx\delta \left \langle r^{2} \right \rangle^{AA'}
\end{eqnarray}

\begin{equation}\label{(7)}
F_{i}=E_{i}f(Z),
\end{equation}
where $E_{i}$, $f(Z)$ are the electronic and relativistic correction factors defined in ref.~\cite{Aufmuth1987} and $C_n$/$C_1$ ratios are tabulated by Seltzer ~\cite{Seltzer1969}.

\begin{figure}
\includegraphics[width = 3.3 in]{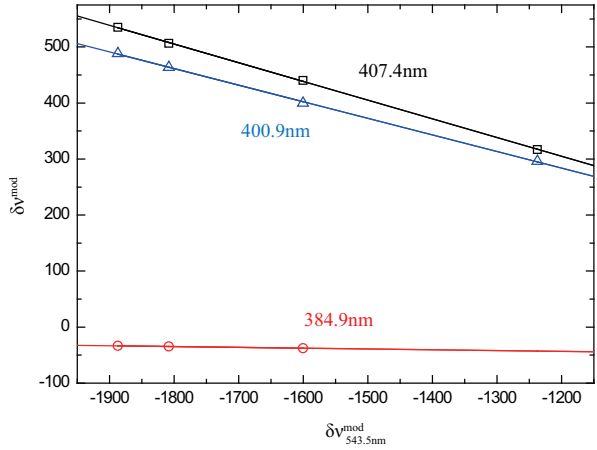}% Here is how to import EPS art
\caption{King plot of the modified isotope shifts of 384.9 nm, 400.9 nm and 407.4 nm transitions with the 543.5 nm transition from W.G. Jin~\cite{Jin1994} as the reference transition. Factor of $\frac{2}{184\times 186}$ was multiplied to both axes for display purposes, which is to get the units in MHz. The $^{180}$W isotope was not measured for 384.9 nm transition. Straight lines passing through the symbols are the least square fit results.  Experimental uncertainties are all within the symbols.}
\label{f:king}
\end{figure}

The NMS can be calculated directly as shown by ref.~\cite{Heilig1974},

\begin{equation}\label{(8)}
M_{i,NMS}=\frac{\nu _{i}}{1836.1},
\end{equation}
where $\nu_i$ is the atomic transition in MHz. However, SMS needs to be extracted from the King plot~\cite{King1984}, which requires isotope shifts from two different transition, one of them being the reference transition with a known SMS. The $ns^2-nsnp$ transitions are often used as a reference transition because its SMS can be evaluated semi-empirically~\cite{Heilig1974}. As we haven't measured any $ns^2-nsnp$ transitions in this work, we used the ($5d^46s^2$ $^5D_1$ - $5d^46s6p$ $^7F_1$) 543.5 nm transition measured from W.G. Jin~\cite{Jin1994} as a reference transition to make the King plots of our measured transitions. These King plots are shown in figure~\ref{f:king} where the axes are the modified isotope shifts of each transitions defined as,

\begin{equation}\label{(9)}
\delta\nu _{i}^{mod}=(\delta\nu _{i}-\delta\nu _{i,NMS})\frac{AA'}{A'-A}.
\end{equation}

From the linear fit, we obtain the relations,

\begin{equation}\label{(10)}
Slope=\frac{E_{i}}{E_{543.5 nm}},
\end{equation}

\begin{equation}\label{(11)}
Intercept=M_{i,SMS}-M_{543.5 nm,SMS}\times \frac{E_{i}}{E_{543.5 nm}}.
\end{equation}

The SMS of 543.5 nm transition is estimated by a semi-empirical relation,

\begin{equation}\label{(12)}
\delta \nu _{543.5 nm,SMS}^{AA'}=(0\pm 0.5)\delta \nu _{543.5 nm,NMS}^{AA'},
\end{equation}
which is only valid for $ns^2-nsnp$ transitions~\cite{Heilig1974}. Combining equations $12\sim14$, we can calculate the SMS of our measured transitions. Knowing NMS and SMS, the FS is readily calculated from eq.~\ref{(3)} and~\ref{(4)}. The extracted values of NMS, SMS and the FS for 384.5 nm, 400.9 nm and 407.4 nm transitions are given in Table ~\ref{table2}. As discuss previously, isotope shifts of the 384.9 nm transition are given by the least square Gaussian fitting method.

\begin{table}
\caption{NMS, SMS, FS, and $\lambda^{AA'}$ are shown for 384.9 nm, 400.9 nm and 407.4 nm Tungsten transitions. 1$\sigma$ errors are shown inside the parentheses. For $\lambda^{AA'}$, we show two independent errors, where the first parenthesis shows the experimental uncertainty from the field shift measurement and the second parenthesis showing theoretical uncertainty of $F_{i}$ from equation~\ref{(5)}.}\label{table2}
\lineup
\begin{tabular}{@{}lllllll}
\br

Transition&Isotopes&$\delta \nu _{i}^{AA'}$&$\delta \nu _{i,NMS}^{AA'}$&$\delta \nu _{i,SMS}^{AA'}$&$\delta \nu _{i,FS}^{AA'}$&$\lambda^{AA'}$\\
(nm)&(A,A')&(MHz)&(MHz)&(MHz)&(MHz)&$(fm^2)$\\
\mr

384.9&184,186&-12.7(1.6)&24.8&-60.2(0.9)&22.6(1.9)&0.0808(68)(57)\\
     &182,184&-10.0(1.6)&25.3&-61.5(0.9)&26.1(1.9)&0.0932(68)(65)\\
     &182,183&\0-4.5(1.6)&12.7&-30.9(0.9)&13.7(1.9)&0.0489(68)(34)\\
&&&&&\\
400.8&184,186&423.6(2.1)&23.8&-71.4(3.7)&471.2(4.3)&0.0810(7)(57)\\
     &182,184&498.4(2.1)&24.3&-72.9(3.8)&547.0(4.3)&0.0929(7)(65)\\
     &182,183&263.2(2.1)&12.2&-36.7(2.1)&287.7(2.9)&0.0489(5)(34)\\
     &180,182&334.0(2.1)&24.9&-74.6(3.9)&383.7(4.4)&0.0651(7)(46)\\
&&&&&\\
407.4&184,186&463.6(2.9)&23.4&-95.5(4.1)&535.7(5.0)&0.0803(7)(56)\\
     &182,184&541.4(2.9)&23.9&-97.6(4.2)&615.1(5.1)&0.0932(8)(65)\\
     &182,183&286.9(2.9)&12.0&-49.1(2.4)&323.9(3.8)&0.0490(6)(34)\\
     &180,182&355.6(2.9)&24.5&-99.8(4.3)&430.9(5.2)&0.0653(8)(46)\\

\br
\end{tabular}
\end{table}

From eq.~\ref{(5)} and~\ref{(7)}, the Field Shift is linked to the nuclear parameter $\lambda^{AA'}$ through $E_i$ and $f(Z)$. The electronic factors $E_{384.9 nm}$, $E_{400.9 nm}$ and $E_{407.4 nm}$ are derived from eq.~\ref{(10)}, with an electronic factor $E_{543.5 nm}=0.40882$ calculated by Aufmuth~\cite{Aufmuth1988a}. The relativistic correction factor $f(Z)$ was calculated by ref.~\cite{Aufmuth1987} using the isotope shift constant of Blundell~\cite{Blundell1985}. The extracted $\lambda^{AA'}$ from three different transitions are shown on the last column of Table~\ref{table2}. All the results show good agreement within the errors. Two independent sources of errors of $\lambda^{AA'}$ are shown inside the parentheses, the first parenthesis showing the experimental uncertainty from the field shift measurement and the second parenthesis showing theoretical uncertainty of $F_{i}$. The theoretical uncertainty in $F_{i}$ comes from fractional error in the process of calculating $E_i$ and $f(Z)$, which is estimated to be $5\%$ for all three transitions~\cite{Heilig1985}. However, the experimental uncertainty of the field shift varies for different transitions.

In order to see the individual error contributions during the process of FS derivation, we combine eq.~\ref{(3)},~\ref{(4)},~\ref{(10)} and~\ref{(11)}, to get,

\begin{equation}\label{(13)}
\delta \nu _{i,FS}^{AA'}=\delta \nu _{i}^{AA'}-\delta \nu _{i,NMS}^{AA'}- \frac{AA'}{A'-A}\times Intercept-\delta \nu _{543.5 nm,SMS}^{AA'}\times Slope.
\end{equation}

\begin{table}
\caption{The sources of uncertainties in $\delta \nu _{i,FS}^{184,186}$ from different transitions of tungsten.}\label{table3}
\begin{indented}
\lineup
\item[]\begin{tabular}{@{}llll}
\br
Source of Uncertainty&\centre{3}{Transition (nm)}\\
\ns
&\crule{3}\\
&384.9&400.9&407.4\\
\mr
Statistical Uncertainty of $\nu _{i}^{184,186}$ (MHz)&1.6&1.6&1.6\\
Systematic Uncertainty of $\nu _{i}^{184,186}$ (MHz)&0.06&1.2&2.4\\
Uncertainty of $\frac{AA'}{A'-A}\times$ Intercept of King plot (MHz)&0.9&1.1&1.4\\
Uncertainty of $\nu _{543.5 nm,SMS}^{184,186}\times$ Slope of King Plot (MHz)&0.2&3.6&3.9\\
\br
\end{tabular}
\end{indented}
\end{table}

We analyze different sources of uncertainties in eq.~\ref{(13)} for all three transitions, which are shown on Table~\ref{table3} for the case of the isotope shift between $^{184}$W and $^{186}$W. Note that the second term of the right hand side of eq.~\ref{(13)}, is a well defined value from eq.~\ref{(8)}, therefore has no error. Row 1 through 3 of Table~\ref{table3} shows uncertainties that are linked to experimental uncertainty of our frequency measurements. However the uncertainty described on the last row, which contains the SMS of the reference line, only rely on the semi-empirical relation of eq.~\ref{(12)}. This becomes the dominating error contribution in the case of 400.9 nm and 407.4 nm transition. Therefore, improving the experimental errors alone, which only affects first three rows of Table~\ref{table3}, would not further decrease the uncertainty of the FS of 400.9 nm and 407.4 nm transitions. Better estimation of the SMS is required for further improvement.

Since the extracted $\lambda^{AA'}$ in Table~\ref{table2} showed good agreement among different transitions, we take the weighted mean of these to report the final values of $\lambda^{AA'}$ and compare with the previous results of ref.~\cite{Aufmuth1987} on Table~\ref{table4}. The errors are presented in the same manner as before, showing the experimental uncertainty inside the first parenthesis and the theoretical uncertainty inside the second parenthesis. We notice experimental uncertainties being much smaller than the theoretical uncertainties for all isotope pairs.

From the above discussion, it is clear that the limiting factors for high precision $\lambda^{AA'}$ measurement mainly comes from the uncertainty in estimation of Specific Mass Shift through semi-empirical relation, and the theoretical uncertainties in the Field Shift. First we discuss about the possibility of SMS calculation. Instead of using the King plot approach, there has been several attempts of multi-configuration calculation of SMS~\cite{Martensson1982,Fonseca1983,Chambaud1984}, however, as pointed out by Aufmuth~\cite{Aufmuth1987}, reliable prediction was never made. Following the approach of ref.~\cite{Chambaud1984}, the specific mass effect perturbs the energy of the electronic state $\psi$ between mass A and A' by an amount of,

\begin{equation}\label{(14)}
\Delta E_{SMS}(\psi ) = \frac{A'-A}{AA'}\left \langle \psi \left | \sum_{i>j}\mathbf{p_{\mathit{i}}}\cdot \mathbf{p_{\mathit{j}}} \right | \psi \right \rangle,
\end{equation}
where $\mathbf{p_{\mathit{i}}}$ is the momentum of the $i^{th}$ electron. Thus, the frequency difference in upper state of $\psi_u$ and lower state of $\psi_l$ can be written as,

\begin{equation}\label{(15)}
\delta \nu _{u\to l,SMS}^{AA'}=\frac{\Delta E_{SMS}(\psi_u )-\Delta E_{SMS}(\psi_l )}{h}.
\end{equation}

%We calculate the SMS of 3 transitions studied in this work by using the relativistic electronic wavefuctions of $\psi_{^7S_3}$
%, $\psi_{^7P_3}$, $\psi_{^7P_4}$, and $\psi_{^5F_1}$. [Here is where we put Titov's calculation and compare with our estimation of SMS from King plot method.]

This would be a more general way of estimating the SMS as it does not rely on the semi-empirical relation shown on eq.~\ref{(12)}, however, the knowledge on tungsten electron wavefunctions are required.

Similarly, we can write down an expression for the field effect perturbing the energy of $\psi$ between mass A and A' by,

\begin{equation}\label{(16)}
\Delta E_{FS}(\psi)=\frac{\pi a_{0}^{3}\left | \psi (0)^{2} \right |}{Z}f(Z)\lambda ^{AA'},
\end{equation}
where $\left | \psi (0)^{2} \right |$ is the non-relativistic electron charge density at the nucleus and $f(Z)$ is the same relativistic correction factor which was mentioned above. The frequency difference between the upper state of $\psi_u$ and lower state of $\psi_l$ becomes,

\begin{equation}\label{(17)}
\delta \nu _{u\to l,FS}^{AA'}=\frac{\Delta E_{FS}(\psi_u )-\Delta E_{FS}(\psi_l )}{h}.
\end{equation}

As indicated from eq.~\ref{(16)}, the FS analysis could be used as a study of $\left | \psi (0)^{2} \right |$. This term could be used for calculation of electric fields inside the atoms, which are important in electron Electric Dipole Moment experiments~\cite{Sandars1965}.

%[There might be a way to calculate relativistic $\left | \psi (0)^{2} \right |$ directly for $\psi_{^7S_3}$
%, $\psi_{^7P_3}$, $\psi_{^7P_4}$, and $\psi_{^5F_1}$. If this is possible from Titov with lower uncertainties than the current calculations, we can reduce the %error of $\lambda ^{AA'}$.]

The tungsten electronic wavefunctions have not been studied in detail so far, which makes the SMS and FS calculations described in eq.~\ref{(14)} $\sim$ eq.~\ref{(17)} unavailable. At the current state, the best way to reduce the fractional error of the nuclear parameter is by defining a relative $\lambda^{AA'}$ as,

\begin{equation}\label{18}
\lambda_{rel}^{AA'}=\frac{\lambda^{AA'}}{\lambda^{184,186}}.
\end{equation}

The fractional error of $\lambda_{rel}^{AA'}$ is much smaller than $\lambda^{AA'}$, which can be seen on the last two columns of Table~\ref{table4}. The results are in very good agreement with ref.~\cite{Aufmuth1987}, with $3\sim16$ times better precision.

\begin{table}
\caption{$\lambda^{AA'}$ and $\lambda^{AA'}_{rel}$ compared with previous results from ref.~\cite{Aufmuth1987}. 1$\sigma$ errors are shown inside the parentheses. For our results, the first parenthesis shows the experimental uncertainty from the field shift measurement and the second parenthesis showing theoretical uncertainty of $E_{i}$ and $f(Z)$ from equation~\ref{(7)}.}\label{table4}
\begin{indented}
\lineup
\item[]\begin{tabular}{@{}lllll}
\br
&\centre{2}{Weighted $\lambda^{AA'}$ $(fm^2)$}&\centre{2}{Weighted $\lambda^{AA'}_{rel}$}\\
\ns
Isotopes&\crule{2}&\crule{2}\\
(A,A')&This work&Ref.~\cite{Aufmuth1987}&This work&Ref.~\cite{Aufmuth1987}\\
\mr
184,186&0.0807$(5)(33)$&0.084(7)&1&1\\
182,184&0.0931$(5)(38)$&0.097(8)&1.1537(11)&1.154(4)\\
182,183&0.0488$(4)(20)$&0.051(5)&0.6047(22)&0.607(5)\\
180,182&0.0652$(5)(26)$&0.068(8)&0.8079(14)&0.808(23)\\
\br
\end{tabular}
\end{indented}
\end{table}

\section{Conclusion}

We performed Laser Induced Fluorescence spectroscopy on a tungsten atomic beam to study isotope shifts and hyperfine structure of three different optical transitions. Three new magnetic hyperfine constants were shown, which agreed with Wyart's configuration of $5d^46s6p$ for $^7P_3$, $^7P_4$ states, and $5d^36s^26p$ for the $^5F_1$ state. The isotope shifts of 384.9 nm, 400.9 nm, and 407.4 nm transitions were analyzed to give nuclear parameters $\lambda$ and $\lambda_{rel}$. We have shown the dominating error contributions in the process of Field Shift and the $\lambda$ determination. The $\lambda_{rel}$, which is an alternative way of representing the nuclear parameter, had smaller fractional error than the absolute $\lambda$. Both cases showed good agreement with the previous studies as well as improved precision. The 400.9 nm transition is related to the spectroscopic diagnostics of tungsten influx rate at ITER.

\section{Appendix : Wavelength Meter Calibration}

\begin{figure}
\includegraphics[width = 3.3 in]{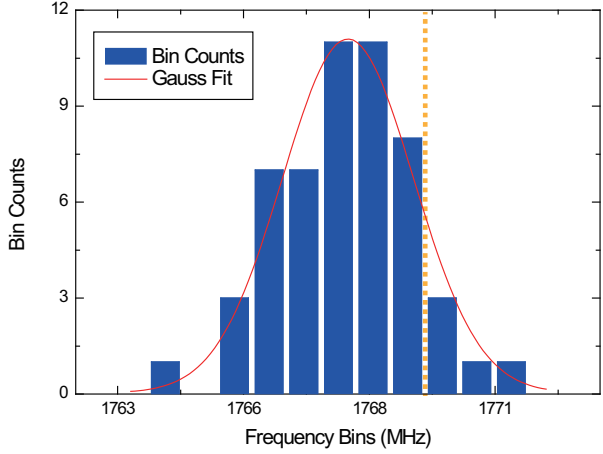}
\caption{The histogram of relative frequency shift between $^{183}$W$\left(a\right)$ and $^{183}$W$\left(b\right)$ of $^7S_3$ $\rightarrow$ $^7P_4$ transition is shown for 53 independent measurements. The Gauss fit for the histogram is shown in red solid line and the reference line position is shown in yellow dashed line. The uncertainty of the reference measurement is $42$kHz which is within the thickness of the dashed line.}
\label{f:histogram}
\end{figure}

The uncertainty in relative frequency measurement could be divided into statistical uncertainty and systematic uncertainty. First we describe the way we characterize the statistical uncertainty. We have measured one of the hyperfine splitting of Tungsten transition multiple times and compare the results with the previously known number. Figure~\ref{f:histogram} shows a histogram of relative frequency shift measurements between $^{183}$W$\left(a\right)$ and $^{183}$W$\left(b\right)$ of $^7S_3$ $\rightarrow$ $^7P_4$ transition, which essentially gives the hyperfine splitting of the $^7S_3$ state as shown on fig.~\ref{f:W401line}. The results were compared with the previous measurements from ref.~\cite{Buttgenbach1979}. The width of the histogram corresponds to the statistical uncertainty of the measurement, which was given as $1.6$MHz from the fit.

In order to test for a possible systematic uncertainty in relative frequency measurements from the wavelength meter, we measured the known Ytterbium(Yb) isotope shifts. Ytterbium was chosen due to a notable high precision isotope shift measurements from Natarajan group~\cite{Pandey2009}.

Having relatively low melting point, Yb atomic beam was produced by conventional method of resistively heated oven, rather than the technique we described in section 2. Multiple apertures along the beam line minimized the Doppler linewidth. A tunable frequency doubled diode laser system (Toptica DL pro) generated 555.6 nm light for the LIF spectroscopy on $^1S_0$ $\rightarrow$ $^3P_1$ transition of Yb. Yb has 7 isotopes, $^{168}$Yb $\sim$ $^{176}$Yb, with 2 of them, $^{171}$Yb and $^{173}$Yb, having hyperfine structure. Detailed analysis on Yb spectrum is omitted, as it goes outside the subject of this section.

%As for estimation of statistical uncertainty of relative frequency measurements, we choose to measure the shift between $^{173}Yb \left (7/2\rightarrow 5/2  \right )$ and $^{171}Yb \left (1/2\rightarrow 1/2  \right )$ multiple times and compare with the measurements from ref.~\cite{Pandey2009}. Figure~\ref{f:histogram} shows a histogram of frequency shift measurements between $^{173}Yb \left (7/2\rightarrow 5/2  \right )$ and $^{171}Yb \left (1/2\rightarrow 1/2  \right )$, along with the reference measurement from ref.~\cite{Pandey2009}. The width of the histogram corresponds to the statistical uncertainty of the measurement, which was given as $1.7MHz$ from the fit.

\begin{table}
\caption{Our measurements of Yb isotopes and hyperfine structure are shown as relative shift from $^{176}$Yb and compared with ref.~\cite{Pandey2009}. The 1 $\sigma$ errors are shown inside the parentheses. As for our results, we show two separate sources of errors, first parenthesis showing the statistical uncertainties and the second parenthesis showing the systematic uncertainties in frequency measurements.}\label{table0}
\begin{indented}
\lineup
\item[]\begin{tabular}{@{}lll}
\br
&\centre{2}{Shift from $^{176}$Yb (MHz)}\\
\ns
Transition&\crule{2}\\
        &This work&Ref.~\cite{Pandey2009}\\
\mr
$^{173}$Yb$_{\left(7/2\rightarrow5/2\right)}$&-1426.3(1.6)(3.6)&-1431.872(60)\\
$^{171}$Yb$_{\left(1/2\rightarrow1/2\right)}$&-1170.6(1.6)(3.6)&-1177.231(60)\\
$^{176}$Yb                                 &\0\0\0\00\0(1.6)(3.6)&\0\0\0\00\0\0\0(60)\\
$^{174}$Yb                                 &\0\0955.2(1.6)(3.6)&\0\0954.832(60)  \\
$^{172}$Yb                                 &\01952.6(1.6)(3.6)&\01954.852(60) \\
$^{170}$Yb                                 &\03235.5(1.6)(3.6)&\03241.177(60) \\
$^{173}$Yb$_{\left(5/2\rightarrow5/2\right)}$&\03258.8(1.6)(3.6)&\03266.243(60) \\
$^{168}$Yb                                 &\04613.4(1.6)(3.6)&\04609.960(60) \\
$^{171}$Yb$_{\left(3/2\rightarrow1/2\right)}$&\04758.3(1.6)(3.6)&\04759.440(60) \\
$^{173}$Yb$_{\left(5/2\rightarrow3/2\right)}$&\04762.4(1.6)(3.6)&\04762.110(60) \\
\br
\end{tabular}
\end{indented}
\end{table}

\begin{figure}
\includegraphics[width = 3.3 in]{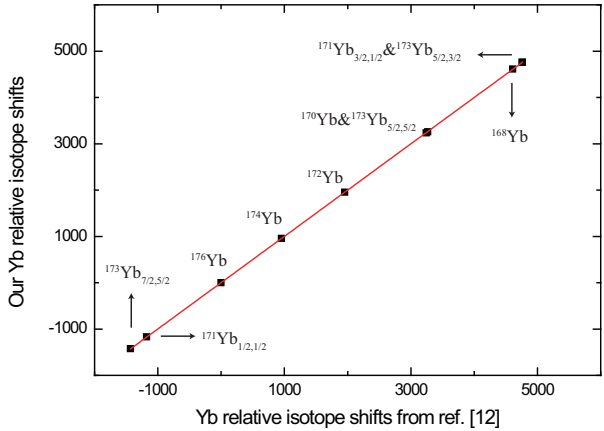}
\caption{Our relative frequency measurements of Yb isotopes and hyperfine structure are shown in black squares. The uncertainty of both ours and the reference measurements are within the symbols. The red solid line shows the least square linear fit of the data points.}
\label{f:YbRefPlot}
\end{figure}

We compare the center line positions of all of our Yb isotopes and hyperfine structure measurements with ref.~\cite{Pandey2009}, which are shown in Table~\ref{table0}. Figure~\ref{f:YbRefPlot} shows the plot of Table~\ref{table0} in black squares, as well as the red solid line showing the least square linear fit. Ideally, this line should have slope of 1, and intercepting at 0. Our fit gave a slope of $0.9990(6)$, and the intercept at $2.0(1.7)$MHz with 1 $\sigma$ errors inside the parenthesis. The error of the slope, which is $6\times10^{-4}$, represents the fractional uncertainty over the measured frequency range. Accordingly, we can assign systematic uncertainty of $3.6$MHz over the range of $6$GHz in the case of Yb isotope shift measurements at 555.6 nm transition. By assuming that the wavelength meter would perform in a similar way at the wavelength range of 385 nm - 410 nm, we can apply these results for the estimation of systematic uncertainty of tungsten transitions being studied in this work.

In conclusion, we report $1.6$MHz of statistical uncertainty and $6\times10^{-4}$ of fractional systematic uncertainty for the frequency measurements from our wavelength meter. The fractional systematic uncertainty converts to 60kHz for 384.9 nm transition, 1.2MHz for 400.9 nm transition, and 2.4MHz for 407.4 nm transition. Adding statistical and systematic uncertainties in quadrature, the total uncertainty of frequency measurements are, 1.6MHz for 384.9 nm transition, 2.1MHz for 400.9 nm transition, and 2.9MHz for 407.4 nm transition.

\section*{References}

\bibliographystyle{unsrt}
\bibliography{JLeeTungsten}

\end{document}